\documentclass[twocolumn,secnumarabic,amssymb, nobibnotes, aps, prd, showpacs]{revtex4}

\usepackage{amsmath,amsthm,amssymb,ulem}
\usepackage[dvips]{graphicx,color}
\begin{document}

\title{Empirical analysis of collective human behavior for extraordinary events in blogosphere}
\author{Yukie Sano$^1$}\email[E-mail: ]{sano.yukie@nihon-u.ac.jp}
\author{Kenta Yamada$^2$}
\author{Hayafumi Watanabe$^3$}
\author{Hideki Takayasu$^{4,5}$}
\author{Misako Takayasu$^3$}

\affiliation{$^1$College of Science and Technology, Nihon University, 7-24-1 Narashinodai, Funabashi, Chiba 274-8501, Japan}
\affiliation{$^2$Waseda Institute for Advanced Study, 1-6-1 Nishi Waseda, Shinjuku-ku, Tokyo 169-8050, Japan}
\affiliation{$^3$Department of Computational Intelligence and Systems Science, Interdisciplinary Graduate School of Science and Engineering, Tokyo Institute of Technology, 4259 Nagatsuta-cho, Midori-ku, Yokohama 226-8502, Japan}
\affiliation{$^4$Sony Computer Science Laboratories, 3-14-13 Higashi-Gotanda, Shinagawa-ku, Tokyo 141-0022, Japan}
\affiliation{$^5$Meiji Institute for Advanced Study of Mathematical Sciences, 1-1-1 Higashimita, Tama-ku, Kawasaki 214-8571, Japan}

\begin{abstract}
To uncover underlying mechanism of collective human dynamics, we survey more than 1.8 billion blog entries and observe the statistical properties of word appearances. We focus on words that show dynamic growth and decay with a tendency to diverge on a certain day. 
After careful pretreatment and fitting method, we found power laws generally approximate the functional forms of growth and decay with various exponents values between -0.1 and -2.5. 
We also observe news words whose frequency increase suddenly and decay following power laws.
In order to explain these dynamics, we propose a simple model of posting blogs involving a keyword, and its validity is checked directly from the data. 
The model suggests that bloggers are not only responding to the latest number of blogs but also 
suffering deadline pressure from the divergence day. 
Our empirical results can be used for predicting the number of blogs in advance and for estimating the period to return to the normal fluctuation level. 
\end{abstract}


\maketitle

\section{Introduction}\label{Sec:intro}
Collective behavior in human society has attracted considerable interest in the last decade~\cite{Nature,Dezso,Ratkiewicz,Yasseri,Kaski,Watanabe,Yamada,Ueno,Mizuno,Sornette2,Facebook,Sornette1,Alfi,JEIC}. 
Because developments in information technology have enabled the storage of large volumes of high-frequency human activity data. 
For instance, detecting bubbles in stock exchange activities~\cite{Watanabe}, 
modeling dealer behavior using real data in the foreign exchange markets~\cite{Yamada}, 
and the empirical analysis of consumer behavior in supermarkets and convenience stores using purchase history and point of sales (POS) data~\cite{Ueno,Mizuno}. 
Human activity data that is collected from the web, for example, YouTube videos and the social network service Facebook, are analyzed to not only explain basic individual human behavior but also elucidate hidden network structures in the society~\cite{Sornette2,Facebook}. 

Here we also use the data from the web to uncover non-trivial mechanism of collective human activities.  
Because word frequency on the web is expected to immediately reflect the real social mood, it has attracted increasing attention among many academic and industrial researchers. In fact, they are stored electronically and analyzed widely. For example, the Library of Congress in the United States which is the largest library in the world has been archiving the entire public tweet of Twitter, a micro-blogging system since 2007, (http://blog.twitter.com/2010/04/tweet-preservation.html). 

A blog is a type of website that is maintained by an individual with entries displayed chronologically with time stamps. 
The term ``blog''  originated from the combination of ``web'' and ``log,'' and was popularized around the year 2000 when free blog services  began to be provided by internet service companies. 
A ``blogger,'' who is an owner of a blog site, can easily upload his/her ``entries'' any time, 
and readers can easily post comments on the blog page.
This interactive quality has contributed to the success of blogs; they are now widely used as 
basic social communication tools. 
The whole blog community is often called the ``blogosphere'', and its scientific study is expected to be a promising new field of science as huge amount of records are compiled as digital data.

In this study, we analyze the keyword appearance rate in blogs in which the functional forms of growth and decay around the peak are approximated by power laws functions of time. 
Similar power laws have been established in other fields of human activity. 
 For example, a power law can describe a decrease in online book sales with an exponent that depends on endogenous or exogenous shocks~\cite{Sornette1}. 
Relaxation in audience number for online movies can also be described by power laws with various values of exponents that reflect the quality of the content~\cite{Sornette2}. 
Alfi et al. found that growth in conference registration numbers is also approximated by the power law diverging at the deadline~\cite{Alfi}.

In Sec.~\ref{Sec:data}, we describe the analyzed data and Japanese blogs. 
In Sec.~\ref{Sec:word}, we introduce our pretreatment procedures and peaked words. 
In Sec.~\ref{Sec:freq}, we focus on the time evolution of these peaked words and prove that they grow and decay with power laws.  
To reproduce power laws, we introduce a simple model of posting blogs in Sec.~\ref{Sec:model}. 
In Sec.~\ref{Sec:predict}, we discuss the predictability of our model from the standpoint of application, and the final section is devoted to conclusions.

\section{Data description}\label{Sec:data}
 The data analyzed in this study was obtained from the blogosphere written in Japanese over a period of four years, 
 from November 1st 2006 to October 31st 2010. 
 \begin{figure}
	\begin{center}
		\includegraphics[width=80mm]{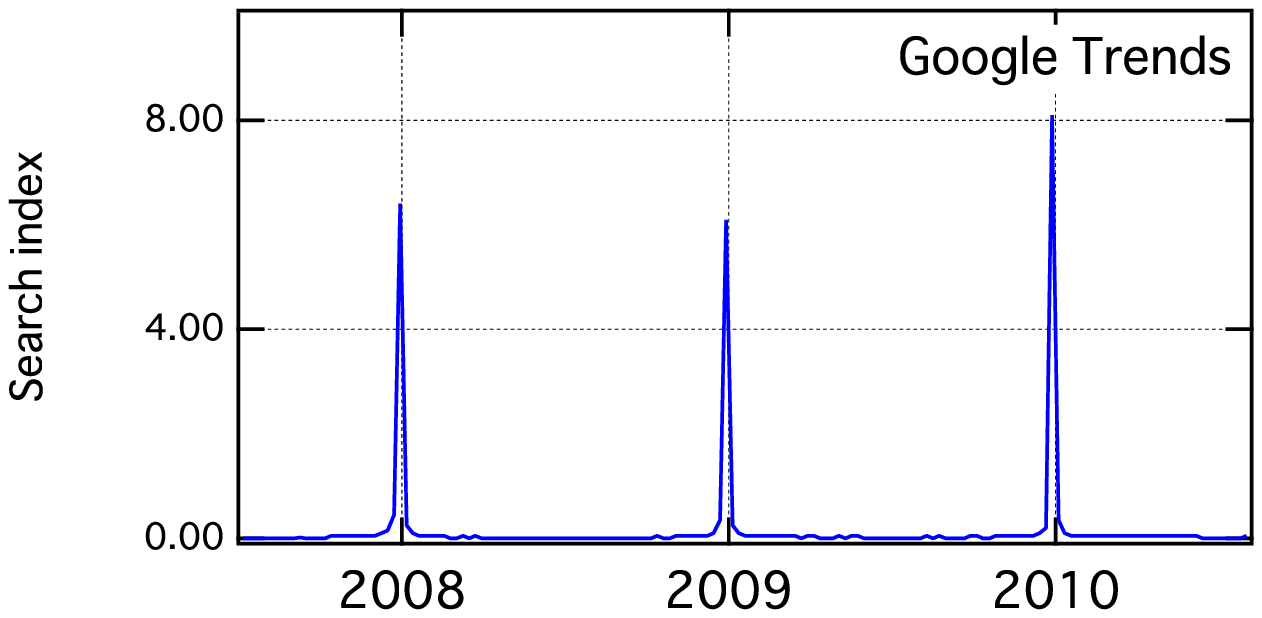}
		\includegraphics[width=80mm]{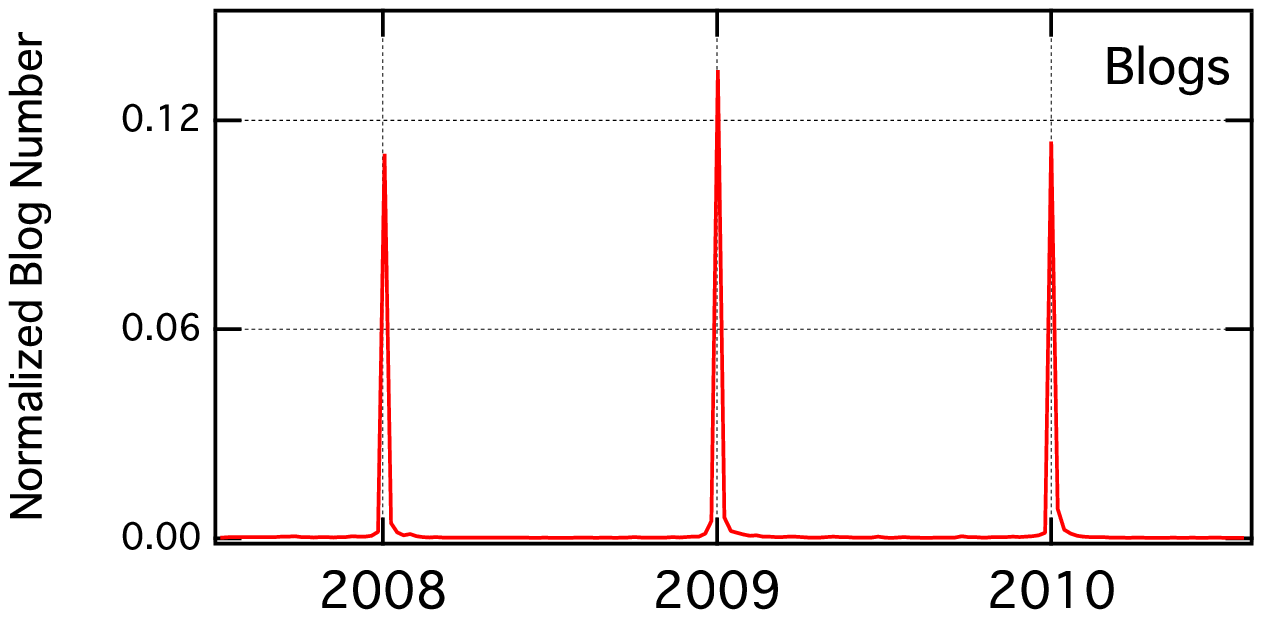}
		\caption{(Color online) Temporal change of the word frequency of  ``April fool'' per week. The results are from Google Trends, which is targeted worldwide and our blog data ``Kuchikomi@kakaricho,'' which is targeted only in Japan . The number of blogs is normalized by the whole number.}
		\label{fig:AprilFool}
	\end{center}
\end{figure}
 According to the technical report by the internet search engine company Technorati (http://technorati.com), that tracked more than 70 million blogs worldwide in 2007, 
the share of Japanese blogs is 37\%, the largest among all languages.  
Although we only analyze the Japanese blogosphere, we show an example in which the dynamic properties in Japanese and English are considerably similar. 
Figure \ref{fig:AprilFool} shows the temporal change of the frequency of the English ``April Fool'' 
observed by Google Trends (http://www.google.com/trends) surveyed worldwide compared to the number of blog entries containing the corresponding Japanese. 
In both cases, we confirm that there is a clear peak on the week including April fools' day. 

In blogosphere research, it is important to note the existence of spam blogs. 
They are automatically generated blogs in which the same words are repeated multiple times, mainly for the purpose of advertising. 
As the share of spams in the Japanese blogosphere is said to be 40\%, it is important to exclude spams from the data. 
We used a new internet service called ``Kuchikomi@kakaricho'' (http://kakaricho.jp) to collect the data. 
This service provides an application programming interface (API) that counts 
the number of entries in which a given target word appeared in a given period by using a search engine technology with a spam filter. 
There are three levels of spam filtering and we apply the middle level, which is known to remove most of the spams while keeping most of the human blogs untouched. 
The API counts the number of entries in the blogs such that if one entry includes the target word multiple times, the word is counted only once. @

The API started crawling the blogosphere on November 1st 2006 and covered major blog service providers. 
It covers more than 1.8 billion blog entries in 15 million blogs accounting for 90\% of the Japanese blogosphere. 

For analysis of Japanese we introduced a pretreatment to separate Japanese words that are not separated by spaces.  
Here we use the commonly used Japanese morphological analyzer ``MeCab'' (http://mecab.sourceforge.net/) to individually separate words according to a dictionary.  
By adding words to its dictionary,  this software can treat multi-word phrases  such as ``April Fool''  as one word, ``April-Fool''. Most of the words used in this study are already listed in the software's dictionary as one word, except names of people. 

\section{Peaked Words}\label{Sec:word}
In the blogosphere, there are special words whose frequency grows or decays around a peak day such as ``April Fool'' with the peak on April 1st. 
In the following discussion, we denote these words as ``peaked words'' and analyze their functional forms of growth and decay. 

\subsection{Pretreatment}\label{Subsec:pre}
\begin{figure}[h]
		\begin{center}
			\includegraphics[width=80mm]{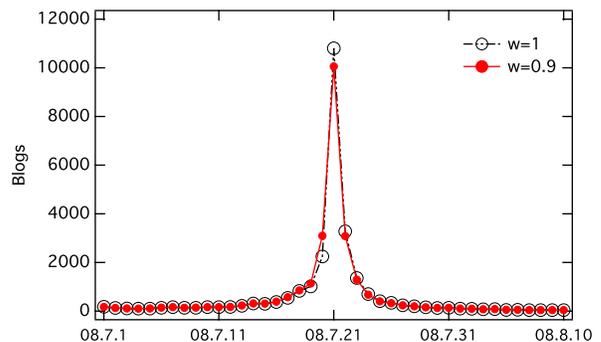}
			\caption{(Color online) Typical example of time series of peaked word ``Marine Day'' in 2008. $w=1$ corresponds to no revision and $w=0.9$ corresponds to modified time series introduced in Sec.~\ref{Subsec:pre}.  Because of the circadian effect, the data of the day after the peak is always higher than that before the peak without modification.}
		\label{fig:Umi}
	\end{center}
\end{figure}

We first apply the following pretreatment to the data to exclude both trivial circadian human activity patterns and systematic noises. 
In this subsection, we mainly focus on statistics of blogosphere itself, not peaked words. 
\begin{description}
\item[Time-Shift]
In blogosphere, although a day starts at 00:00:00, there are many bloggers who are active at midnight. 
Therefore, we examine the complete circadian activity pattern and introduce a type of correction pretreatment for our daily data. 
For this purpose, we randomly chose the data of 10,000 bloggers with the details of their activities time stamped in seconds. 
By counting the number of entries posted at every hour, a circadian activity pattern is plotted in Fig.~\ref{fig:Hours}. 
The solid line shows the 24-hour-activity pattern obtained directly from the data. However, 
we discovered there are a certain number of blog entries with time stamps that are exactly 00:00:00. 
We consider this time stamp to be caused by an artificial systematic spec or error,  
and we exclude this data from the statistics when capturing the circadian pattern. 
The red bars in Fig.~\ref{fig:Hours} show the revised circadian activity pattern. 
Using a 24-hours clock, we find that blogging activity is lowest around 4:00, and thus we consider 
the start of a day at 5:00 to be reasonable. 
Because the share of activity in the interval between 0:00 and 5:00 is approximately 10\% of the complete activity 
of a day, we can correct the daily number of blog entries including the $j$-th target word at the 
$t$-th day $\tilde{x_j}(t)$ by the following equation
\begin{equation}
	x'_j(t)=w\tilde{x_j}(t) + (1-w)\tilde{x_j}(t+1), 
	\label{eq:ts}
\end{equation}
where the weight is set as $w=0.9$. 
With this modification, we can determine the time-shifted time series. 
In Fig. \ref{fig:Umi}, open circles show the original daily data in which $w=1.0$ in Eq.~(\ref{eq:ts}), and colored circles show time-shifted data in which $w=0.9$. The time-shifted data shows a more symmetric pattern than the original data. 
 We also apply this procedure to determine the time series of the total number of blog entries per day $x'(t)$.
To clarify the effect of this time-shift procedure, we also show results without this time-shift procedure in Appendix \ref{App:NoShift}. 
\begin{figure}
	\begin{center}
		\includegraphics[width=80mm]{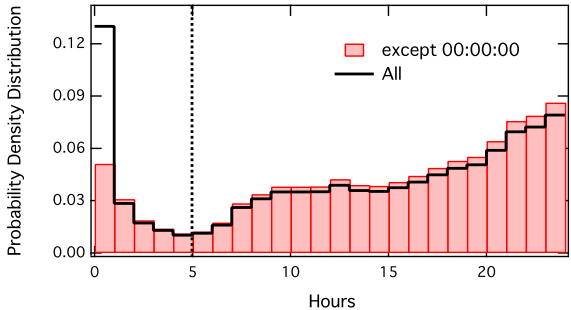}
		\caption{(Color online) Probability density distribution of circadian activity of blog posted by 10000 bloggers. 
		Solid line is calculated from all entries and the red bar is from the entries excepted that have 
		time stamp of 00:00:00. In both, 4:00 is the smallest ratio in a day.}
		\label{fig:Hours}
	\end{center}
\end{figure}

\item[Normalization]
There are non-uniform and non-stationary properties in the total number of entries per day \cite{JEIC}. 
For example, there was a sudden drop in February 2007 that was caused by search engine software's system maintenance. 
In order to reduce the systematic fluctuations caused by such non-uniform properties, we apply the following normalization procedure. 
There is already a method to separate internal and external noises~\cite{Barabasi}, which simply deducts external factor depending on its ratio of the total traffic. They assume that each traffic $x'_j(t)$ in a small component $j \in N$ is consisted of the total traffic $x'(t)$ without overlap, where $\sum_{j=1}^{N}{x'_j(t)}=x'(t)$. 
Here, we simply divide $x'_j(t)$ by the total number, $x'(t)$. 
The normalized number of entries for the $j$-th word on the $t$-th day is defined by  
$x_j(t)=x'_j(t)\frac{\langle x' \rangle}{x'(t)}$, where $\langle x' \rangle$  denotes the mean value of  $x'(t)$ that is averaged over the entire observation period. 
This normalized quantity is proportional to the probability that a blog contains the $j$-th word on the $t$-th day, and it is not necessarily an integer. 

By introducing this normalization, the fluctuations caused by the aforementioned non-uniform properties can be reduced. 
In this study, we measure the word frequency using this normalization procedure. 
\end{description}

\subsection{Word Selection}\label{Subsec:Word}
We determine candidates for peaked words in the following three categories. 

\begin{description}
\item[Event]
We selected the names of 14 public holidays and 16 major annual events in Japan. 
The appearance for these words grows and decays around the date of the event. 
In addition, these are words affiliated with an event, such as ``Santa Claus'' for ``Christmas'' and we can 
observe similar growth and decay behavior for those words. However, in this analysis we neglected such affiliated words. 

\item[Date]
We selected dates such as ``May 9th,'' resulting in 365 words. 
There are many blog entries that announce some special day, e.g., birthday and festival. 
Growth and decay of these words always show a clear peak at the date.

\item[News]
A word such as ``earthquake'' occurs suddenly right after the occurrence of the event and the word appearance rate 
generally decays slowly. 
In order to observe the functional form of such decay after a significant event, 
we selected names of the places impacted by earthquakes. 
We also selected 33 names of famous people who died suddenly. 
In addition, we included the names of the Japanese scientists who received 
a Nobel Prize during our observation period. 
\end{description}

\section{Dynamics of Peaked Words}\label{Sec:freq}
We call the slopes before the peak day, ``fore-slopes'' and those after the peak day, ``after-slopes'', and we examine both in this section. 
As no standard method is known for checking the validity of approximation by a power law time evolution for given time series, we apply a statistical test for power law function introduced by Preis et al.~\cite{Preis} that is based on Kolomogorov-Smirnov statistical test~\cite{Clauset}.

\subsection{Method}
We define the number of days in each slope by the number of consecutive days in which the word frequency is larger than the median value $\bar{x_j}$ from the peak. 
The median value is estimated throughout the entire observation period. 
Then we approximate the functional form of the slopes using two models, a power law and an exponential law. 
\begin{equation}
	x_j(t)-\bar{x_j}=A_j {\left | t_c-t\right|}^{-\alpha_j}
	\label{eq:model}
\end{equation} 
\begin{equation}
	x_j(t)-\bar{x_j}=B_j \exp{\left( -\beta_j \left | t_c-t\right| \right)}
	\label{eq:model2}
\end{equation} 
The parameters of these models, $\alpha_j$, $A_j$, $\beta_j$, and $B_j$, are determined by the least squares method. 
The fitting region is $[t_c\pm1, t_c\pm n]$ where $n$ is the number of days in slope. 
Then we apply the Kolmogorov-Smirnov goodness of fit test, for choosing the better model. 
It was originally used as a statistical test for distributions. Here, we apply it for evaluation of the statistical fitness of the functional form of the time series. 
For both models we calculate the KS statistic $D$, representing the deviation, is defined as
\begin{equation}
	D=\max_{t \in [t_c\pm1, t_c\pm n] } \left | X^{(empirical)}_j(t) - X^{(model)}_j(t) \right|,
	\label{eq:ksd}
\end{equation}
where $ X^{(empirical)}_j(t)$ is the cumulative number of empirical value which is counted from the data, and  $X^{(model)}_j(t)$ is the cumulative number which is calculated from the model. 
In both cases, numbers are normalized by $X_j(t_c\pm1)$. 
By comparing the values of $D$ for both models, the power law model is accepted if the $D$-value for the power law is smaller. 
In the case that the power law is accepted, we check the validity of the model as introduced in \cite{Preis}. 
First we generate 1000 synthetic data set. One data set contains $n$ data points. 
Synthetic data points are generated randomly following the normal distribution with the mean value is best estimated from the model $x^{(model)}_j(t)$ and standard deviation is $\sigma(x^{(model)}_j(t))$ as follows 
\begin{equation}
	\sigma(\langle x_j \rangle) \simeq \sqrt{\langle x_j \rangle \left(1+a^2 \langle x_j \rangle \right)},
	\label{eq:RD}
\end{equation}
where $a=\frac{\sqrt{{\langle X^2 \rangle}_c}}{\langle X \rangle}=0.08$ is a constant parameter characterizing the fluctuation in the number of all bloggers which is determined independently of the word (see Appendix \ref{App:ERD} for theoretical derivation of Eq.~(\ref{eq:RD})). 
For each synthetic time series, we compare its $D$-value with that of the empirical one. 
We count the number of cases in which the $D$-value for the synthetic time series is larger. If the number of such cases are less than 100 out of the 1000 synthetic samples, we accept the power law model as $q=0.1$. 
Contrary to the ordinary sense of $p$-value, the power law hypothesis is considered to be valid for larger $q$.
Thus if the $q$ is close to 1, then the difference between empirical data and the model can be attributed to statistical fluctuation alone and we accept power law hypothesis. If the $q$ is smaller than 0.1, we reject power law hypothesis. 
We change the border of the fitting region $n$ from 5 days to maximum slope length. 
The value of power exponent, $\alpha_j$, is given bye the value for the case with the largest $n$. 

\subsection{Results}\label{Subsec:power}
 \begin{figure}
	\begin{center}
		\includegraphics[width=80mm]{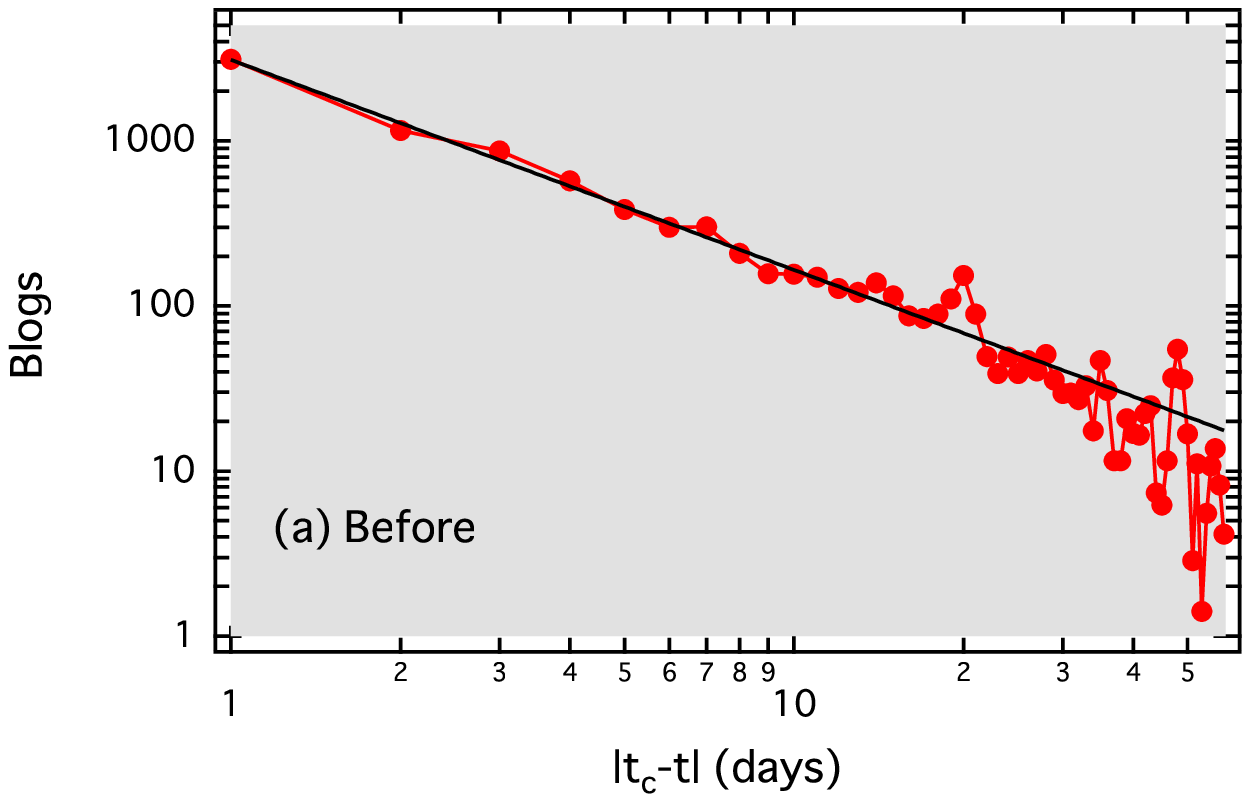}
		\includegraphics[width=80mm]{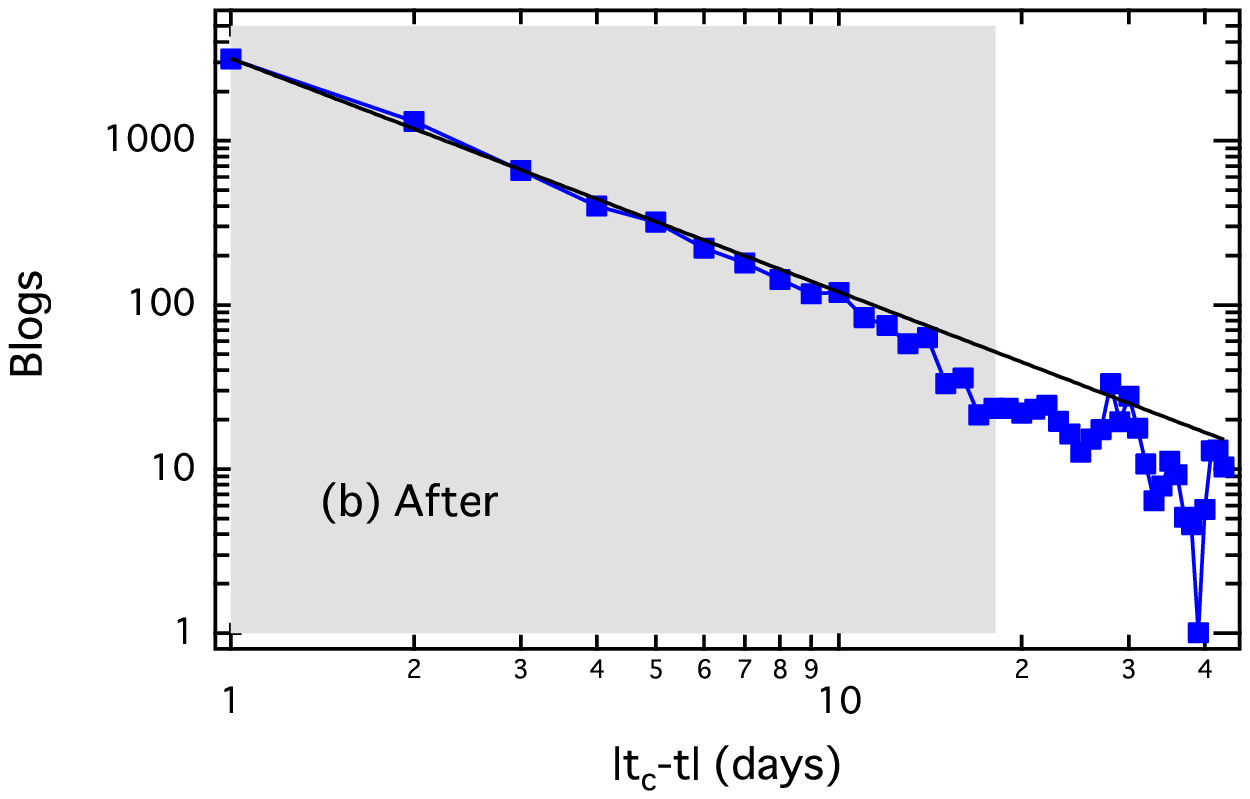}
		\caption{(Color online) Examples of data fitting by power laws of ``Marine Day'' in 2008 for fore-slope (a) and after-slope (b) plotted in log-log scale. For fore-slope, models are fitted by Eq.~(\ref{eq:model}) with $\alpha_j=1.27$ and $A_j=3100$ ($q=0.239$, $n=58$). 
		For after-slope, $\alpha_j=1.42$ and $A_j=3171$ ($q=0.108$, $n=18$). The shaded area shows the interval $n$ in which the power law model is accepted.}
		\label{fig:Fitting}
	\end{center}
\end{figure}
 
\begin{figure}[h]
	\begin{center}
		\includegraphics[width=80mm]{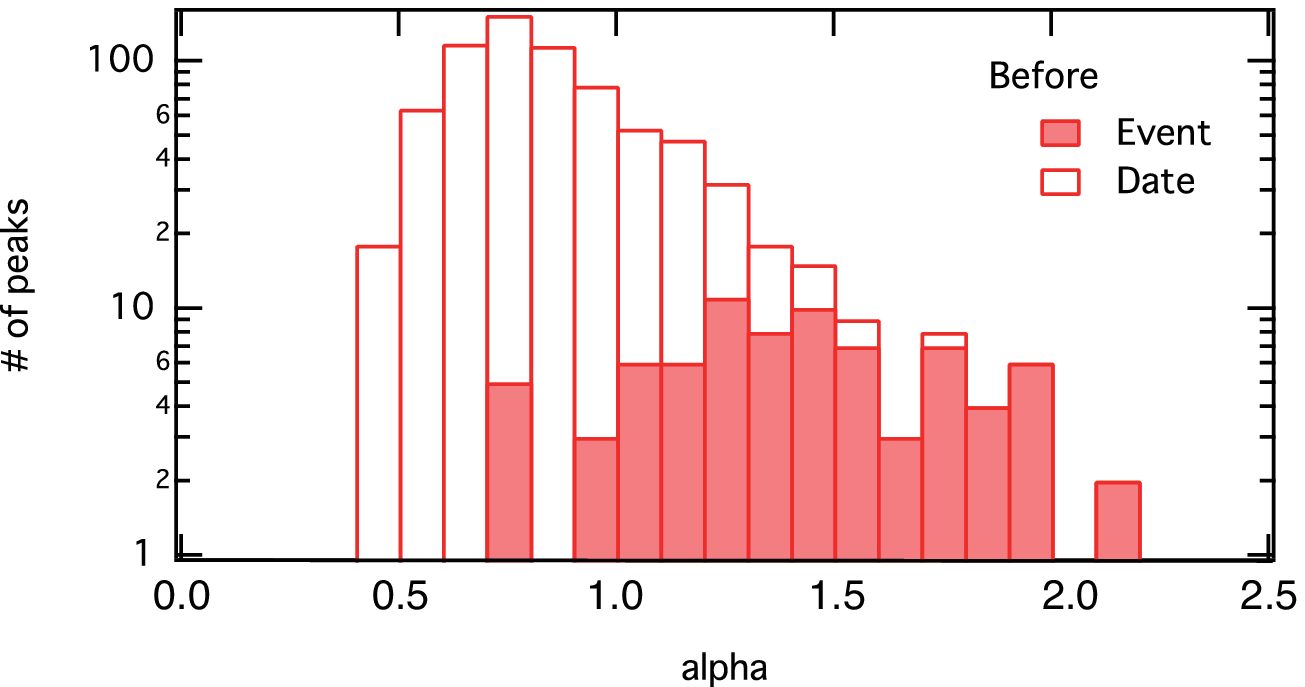}
		\includegraphics[width=80mm]{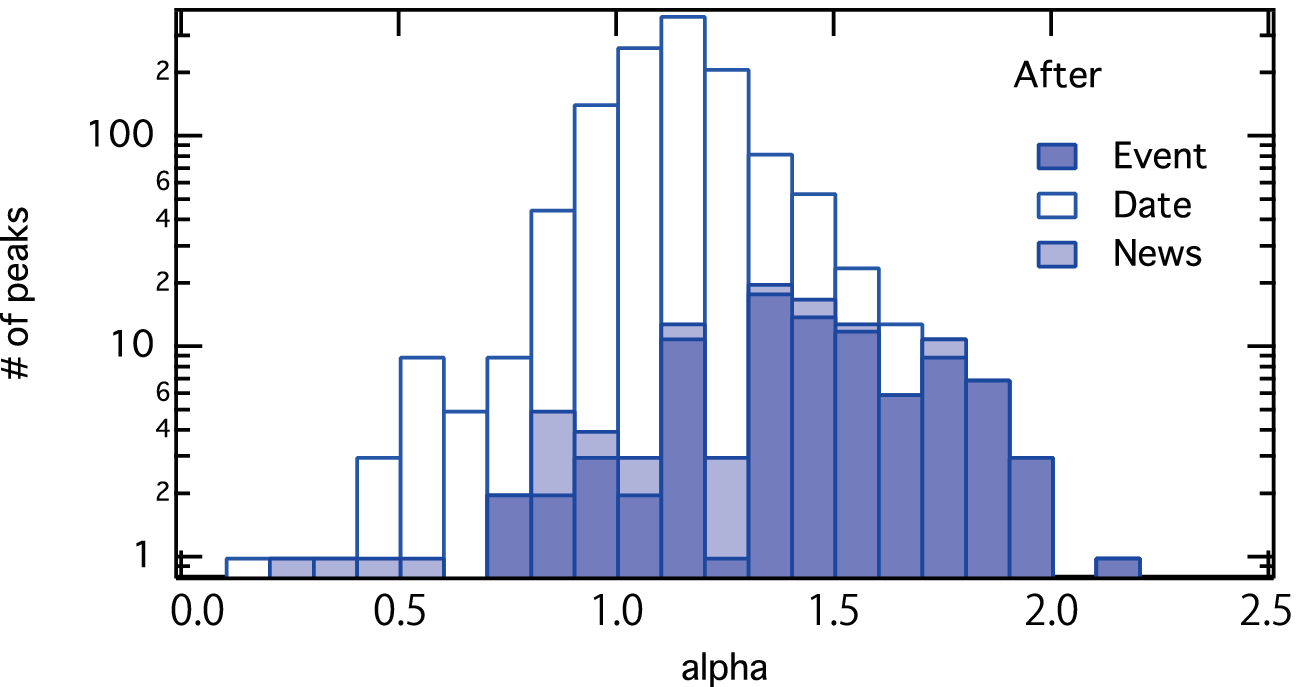}
		\caption{(Color online) Distribution of power exponent $\alpha_j$ of the fore-slopes (a) and the after-slopes (b). 
Mean value of $\alpha_j$ of fore-slope is 0.85 $\pm$ 0.30 and after-slope is 1.13 $\pm$ 0.21. Bars are colored by three categories of peaked words.}
		\label{fig:Alphas}
	\end{center}
\end{figure}

  Figure \ref{fig:Fitting} is a typical result of data fitting for the word ``Marine Day'' in 2008 with log-log scale, as shown in Fig. \ref{fig:Umi} with linear scale. 
For all cases of power law fitting, the distribution of the estimated power exponents are shown in Fig. \ref{fig:Alphas} and summarized in Tab. \ref{tab:Alpha}. 

The absolute value of the power exponents of the after-slopes is larger than that of the fore-slopes in 58\% of the 65 samples for Event, and 80.6\% of the 603 samples for Date. 
For Date, we confirm significant difference between fore-slopes and after-slopes by $t$-test with $p$-value $< 2 \times  10^{-16}$ while it is rejected with $p$-value $=0.80$ for Event.  
The number of days of the after-slopes is larger than that of the fore-slopes in 55\% of the 65 samples for Event, and 65.8\% of the 603 samples for Date. 
For Date, we confirm significant difference between fore-slopes and after-slopes by KS-test with $p$-value $< 2 \times  10^{-16}$ while it is rejected with $p$-value $=0.22$ for Event.  

 In the case of the news words, there is no fore-slope and we cannot compare the values of the exponents before and after the peak. 
 The absolute values of the exponent after the peak tend to be estimated as smaller for high impact news because of the effect of sequential broadcasts after the news. 
 For example, in the case of the sudden death of the world famous entertainer Michael Jackson,
  which marked the peak day, there was a funeral service after a few days and a memorial CD released 
  after a few weeks. 
  Both can be regarded as aftershocks that remind us of the main news. 
  Because of such repetition, the keyword appearance rate after the peak day is enhanced, 
  the decay of the word appearance becomes slower, and the power exponent tends to take a smaller value.
  
 \begin{table}
	\caption{Mean values of power exponent $\alpha_j$ with standard deviations and medians of slope days $n$.}
	\label{tab:Alpha}
	\begin{center}
		\begin{tabular}{ccccc} \hline \hline
			 &  & $\alpha_j$ & $n$ (days) & \# samples\\ \hline
			Event & Before &1.40 $\pm$ 0.38 & 10 & 83\\ 
			 & After &  1.44 $\pm$ 0.28 & 16 & 91 \\ 
			Date & Before & 0.79 $\pm$ 0.38 & 9 & 776 \\ 
			 & After & 1.11 $\pm$ 0.16 & 21  &1229  \\ 
			News & After & 1.09 $\pm$ 0.45  & 10 & 21 \\ \hline
			All & Before & 0.85 $\pm$ 0.30 & 9  & 859  \\
			 & After & 1.13 $\pm$ 0.21 & 20 &1341 \\ \hline  \hline
		\end{tabular}
	\end{center}
\end{table}

\subsection{An extreme case ``Tsunami''}\label{subset:tsunami}
\begin{figure}[h]
	\begin{center}
		\includegraphics[width=80mm]{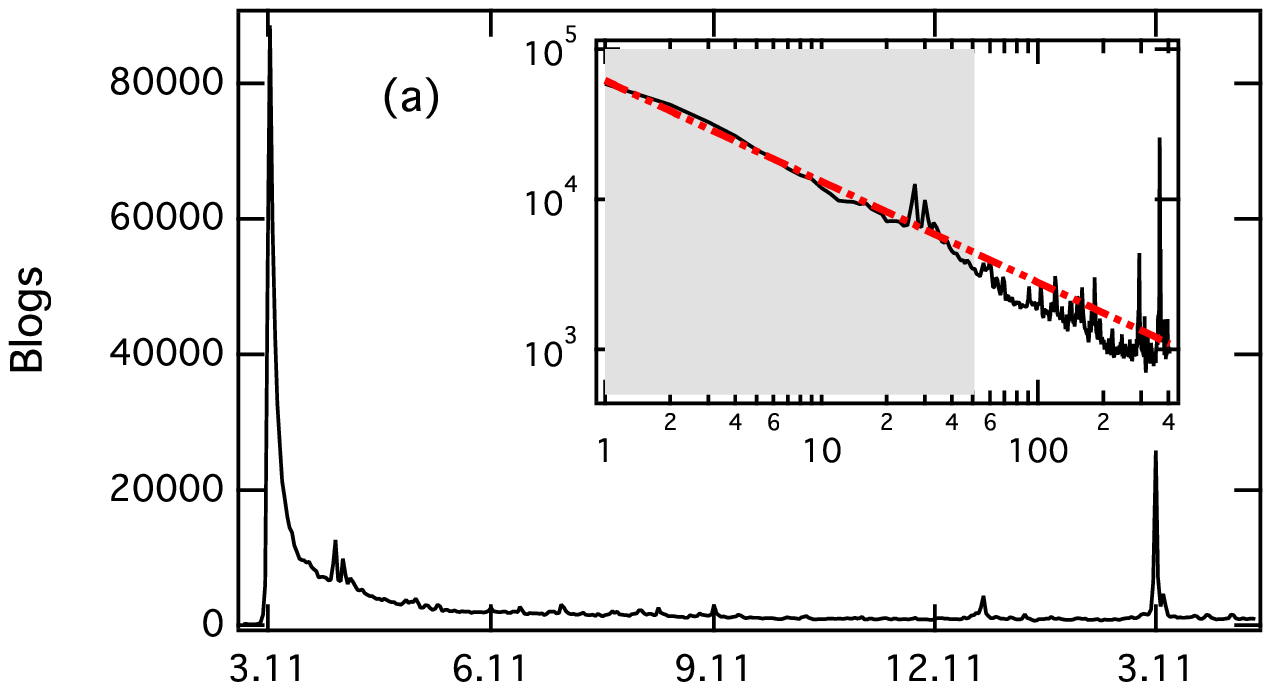}
		\includegraphics[width=80mm]{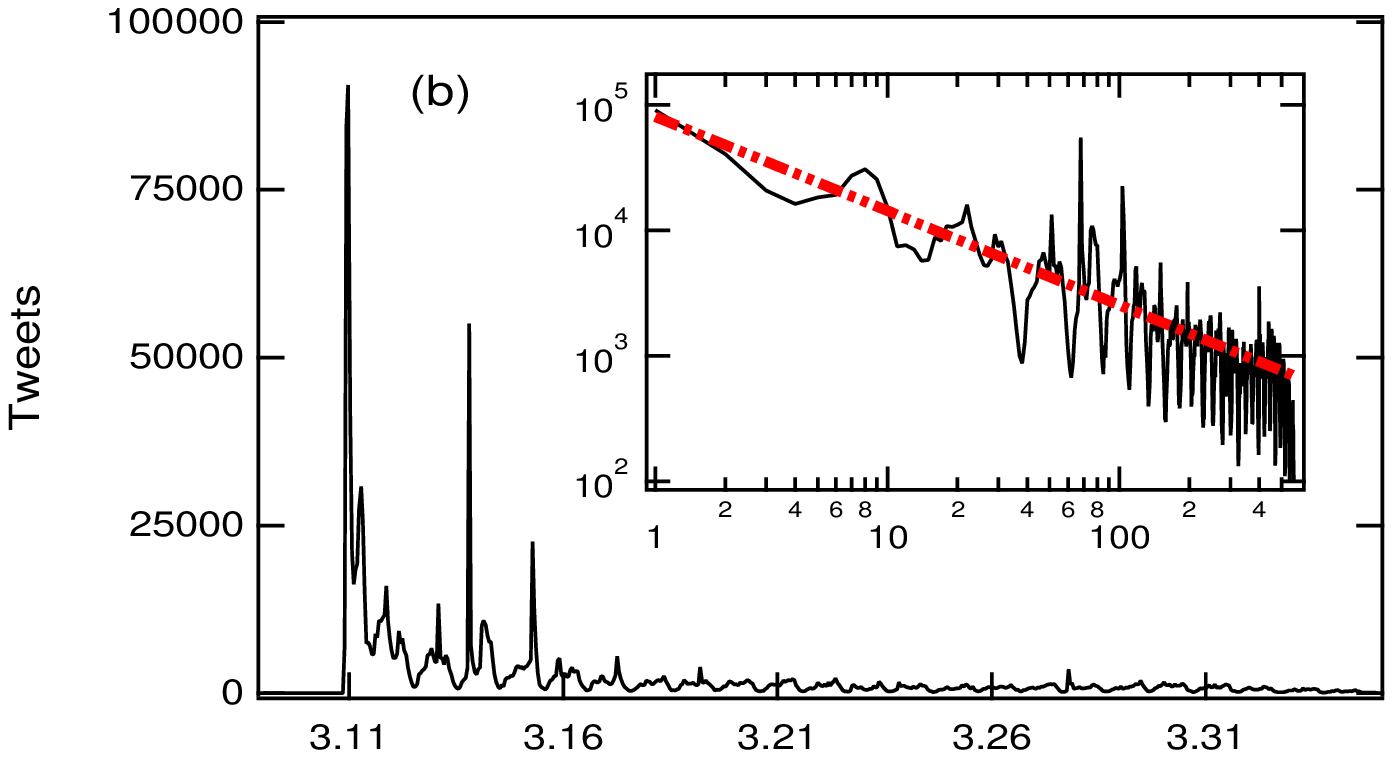}
		\caption{(Color online) Decay of ``Tsunami'' observed for blogs for 417 days since March 1st 2011 (a) and tweets for 648 hours (27 days) since March 9th 2011 (b). Horizontal time step size is per day for blogs and per hour for tweets. (Inset) Log-log plots of the time series. Red dashed lines show the slope of power law with the exponent $\alpha_j=0.67$ for blogs and $\alpha_j=0.75$ for tweets.}
		\label{fig:Tsunami}
	\end{center}
\end{figure}

The power law decay per day of the word ``Tsunami'' in the Japanese blogosphere is shown in Fig.~\ref{fig:Tsunami}(a). 
The peak day was March 12th 2011, the day after the quake with 142617 posts or 12.6 \% of all blog posts in raw data. 
After pretreatment of time-shift and normalization, the estimated power exponent $\alpha_j$ is 0.67 with $A_j=61788$ ($n$=50) using Eq.~(\ref{eq:model}). 
It is expected to take approximately 8623 days ($\sim$ 23.4 years) to return to the normal fluctuation level if we simply broaden power law function. 
The normal fluctuation level was 140 appearances per day, estimated from the data one month before the quake. 
Although most of the news words decay in approximately 10 days, 
the case of ``Tsunami'' is a rare exception because the number of entries is still ten times higher than before the peak even for a year after the quake. 

 Twitter also shows similar power law behavior even though the time resolution is different. 
Figure \ref{fig:Tsunami}(b) shows the number of tweets measured per hour that include ``Tsunami'' that is calculated based on 1397783 tweets. 
We believe that this type of power law reflects the robustness of the empirically observed dynamics of collective human behavior. 

\section{The Model}\label{Sec:model}
In this section, we propose a simple dynamic model to describe the typical power law  growth and decay of frequency of blogs with peaked words. 
There is already a simple model to describe people's universal behavior before a deadline by assuming pressure inversely proportional to the remaining time~\cite{Alfi}. 
As this simple model can describe only the special case $\alpha=1$, a kind of utility function that includes the tendency to postpone the action is introduced to describe the general case. 
Here, we introduce another approach to describe the general case. 
We introduce the following two assumptions for the number changes of blogs including the $j$-th target word at $t$-th day, $\Delta x_j(t)=x_j(t+1)-x_j(t)$, increments for fore-slope and decrements for after-slope.
\begin{enumerate}
	\item The pressure from the peak day $t_c$ works inversely proportional to the time, $1/|t_c-t|$~\cite{Alfi}.  
	\item The number of changes $\Delta x_j(t)$ is proportional to the number of blogs including the $j$-th target word, $x_j(t)$.
\end{enumerate}
We can write these two assumptions into mathematical form in continuous case as we assume $\Delta x_j(t) \simeq\dfrac{\mathrm{d}x_j(t)}{\mathrm{d}t}$. 
The time evolution of blogs for the fore-slope is given as
\begin{equation}
	\dfrac{\mathrm{d}x_j(t)}{\mathrm{d}t} = \alpha_j^{(fore)} \cdot \frac{x_j(t)}{(t_c-t)}+ f(t),
	\label{eq:m1}
\end{equation}
where $f(t)$ is an independent noise with zero mean. The value $\alpha_j^{(fore)}>0$ is a proportionality factor that describes the effect of the above-mentioned two assumptions. Similarly, the decrement of the after-slope is given as
\begin{equation}
	\dfrac{\mathrm{d}x_j(t)}{\mathrm{d}t} = - \alpha_j^{(after)} \cdot \frac{x_j(t)}{(t-t_c)}+ f(t), 
	\label{eq:m2}
\end{equation}
where $\alpha_j^{(after)}>0$ is also a proportionality factor that describes the  effect of the two assumptions. 
Because we know that blogs decrease after $t_c$, we add a negative sign in the Eq.~(\ref{eq:m2}). 
It is easy to confirm that both Eqs.~(\ref{eq:m1}) and (\ref{eq:m2}) derive the power law divergence, Eq.~(\ref{eq:model}), in the case with no noise term $f(t)$. 
Thus, $x_j(t)\propto{(t_c-t)}^{-\alpha_j^{(fore)}}$ for fore-slopes and $x_j(t)\propto{(t-t_c)}^{-\alpha_j^{(after)}}$ for after-slopes. 
In the case that there is no pressure from the peak day $t_c$, blog dynamics follow Eq.~(\ref{eq:model2}) of the exponential law. 

As a check of our assumption, we rewrite Eqs.~(\ref{eq:m1}) and (\ref{eq:m2}) into the following desecrate form without the noise term $f(t)$, and we calculate left-hand-side and right-hand-side values from the real data. 
Note that $t_c$ is not necessarily an integer since the divergence point is expected to exist in single day time period. 
\begin{equation}
	\frac{|\Delta x_j(t)|}{x_j(t)} \simeq \alpha_j \cdot  \frac{1}{|t-t_c|}
	\label{eq:m3}
\end{equation}
\begin{figure}[ht]
\begin{center}
	\includegraphics[width=80mm]{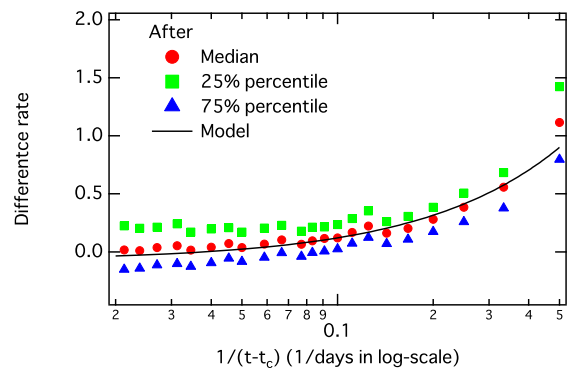}
	\caption{(Color online) Relationship between difference rates $|\Delta x_j(t)|/x_j(t)$ and the absolute inverse number of  time from the peak calculated by Eq.~(\ref{eq:m3}) for after-slopes, which summarized 1341 samples shown in Tab.~\ref{tab:Alpha}. Solid line shows the theoretical relation, Eq.~(\ref{eq:m3}), with the averaged empirical value, $\alpha_j=1.13$. }
	\label{fig:DiffRate}
\end{center}
\end{figure}
We survey all 1341 keywords for after-slopes as listed in Table.~\ref{tab:Alpha}, and the median, upper and lower quantile points are plotted in Fig.~\ref{fig:DiffRate}
As known from this figure, the median and quantile points fit well with the theoretical curve. This means that for majority of words the relation Eq.~(\ref{eq:m3}) holds implying that blog number changes are proportional to the number of recently written blogs, $x_j(t)$, and it is also promotional to $1/(t-t_c)$.

\begin{figure}[ht!!!]
	\begin{center}
		\includegraphics[width=80mm]{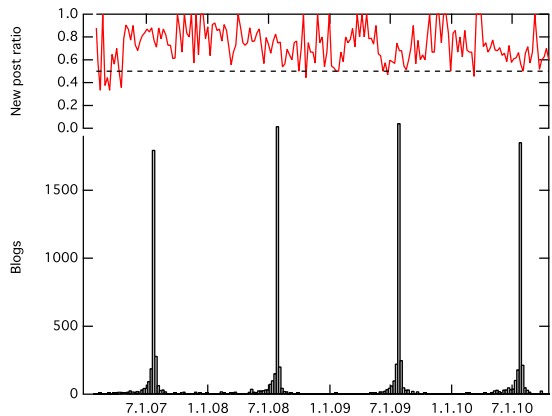}
		\caption{(Color online) Number of blog entries including ``Marine Day'' per week which is calculated from randomly selected 30 thousands bloggers 
		(bottom), and the ratio of the number of newly posted bloggers in the same week (top).  The dashed line shows the ratio 0.5.}
		\label{fig:Newly}
	\end{center}
\end{figure}

Now we know that the above relation Eq.~(\ref{eq:m3}) holds as a whole system, however, 
there remain two scenarios to realize this: one is the case that main bloggers forming the peaked behavior are repeaters and the assumptions hold for each blogger individually, 
and the second case is that the main bloggers are newly joined bloggers and the assumptions holds for general bloggers implying existence of collective interaction in the blogosphere. 
In order to clarify which is the right scenario, we pay attention to randomly chosen 30,000 bloggers whose activities can be traced precisely. 
For all these bloggers we observe the days when they posted the typical keyword ``Marine Day''. 
We count the total number of blogs including this keyword among these bloggers for each week as plotted in Fig.~\ref{fig:Newly} bottom, also we count the number of bloggers who posted the keyword for the first time, and the ratio of the number of new comers over the total number in the week is plotted in Fig.~\ref{fig:Newly} top. 
As known from this figure we confirm that the share of repeaters in the peaked behavior is generally less than half, namely,
 the power law behavior is formed mainly by newly joined bloggers. 
 Similar results are confirmed also for some other typical keywords. This fact implies that the second scenario is correct and the factor $\alpha$ characterizes the strength of influence of written blogs to general bloggers representing the existence of interaction in the blogosphere.

For fore-slopes there is a natural reason of appearance of factor, $1/(t_c-t)$, in Eq.~(\ref{eq:m1}) explained by the deadline effect~\cite{Alfi}, that is, a blogger who plans to post the keyword ``Marine Day'' may think that there are $t_c-t$ days before the deadline and a posting date can be chosen from $t_c-t$ candidates, so the probability of posting a blog on the day is proportional to $1/(t_c-t)$. 
This effect can be regarded as a universal property for each blogger individually. 

On the other hand the reason for after-slopes is less obvious. 
For a blogger who wants to post the keyword after the event, the probability of writing a blog including the keyword might be proportional to the decay of strength of memory. 
In the field of psychology, the functional form of memory decay is usually approximated by a nonlinear function~\cite{Ebbinghaus}, 
and here, as a simplest assumption we introduce the inverse power law of memory decay from the deadline, $1/(t-t_c)$, 
which has the same functional form as the case of fore-slopes. 
With this assumption we can explain the non-trivial exponents of power law behaviors of blogs 
by introducing the factor $\alpha$, that describes the strength of influence of written blogs to general bloggers. 

\section{Predicability of  Frequency}\label{Sec:predict}
\begin{figure}[th]
	\begin{center}
		\includegraphics[width=70mm]{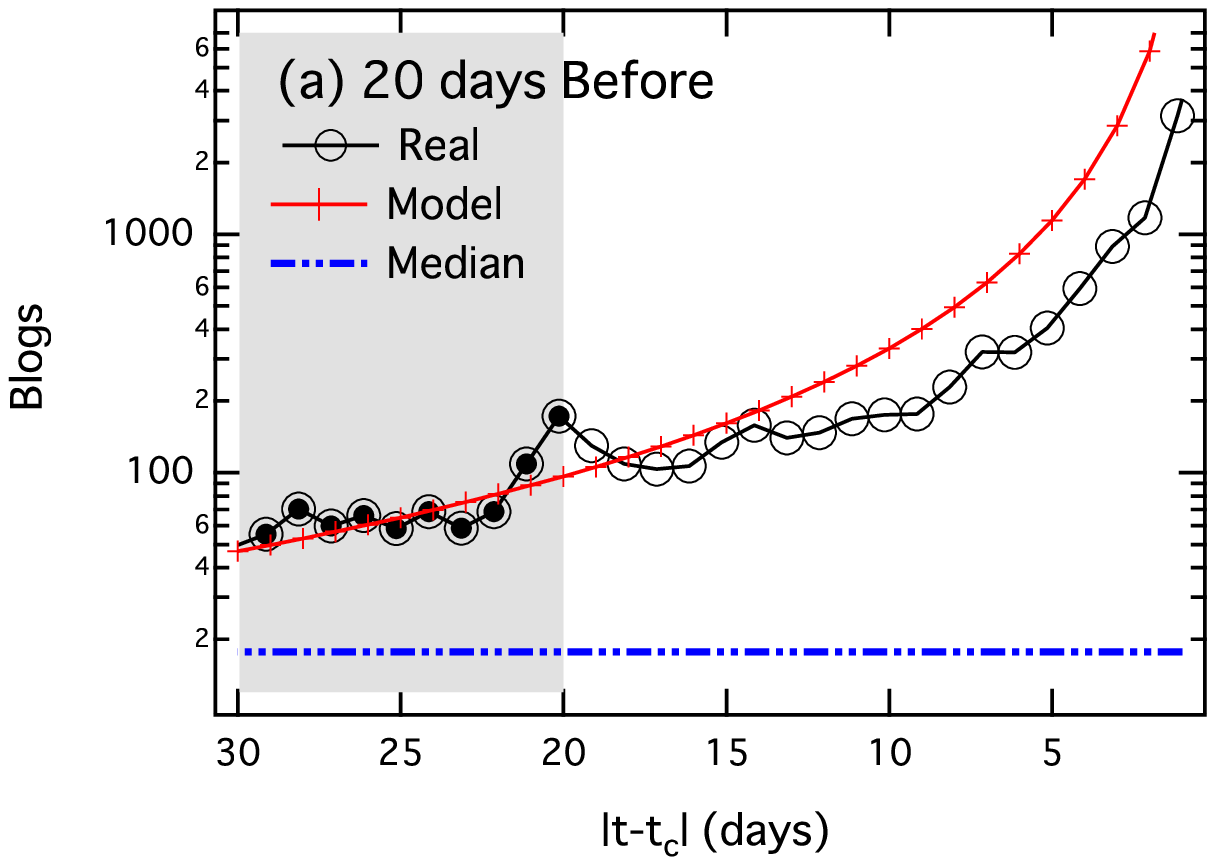}
		\includegraphics[width=70mm]{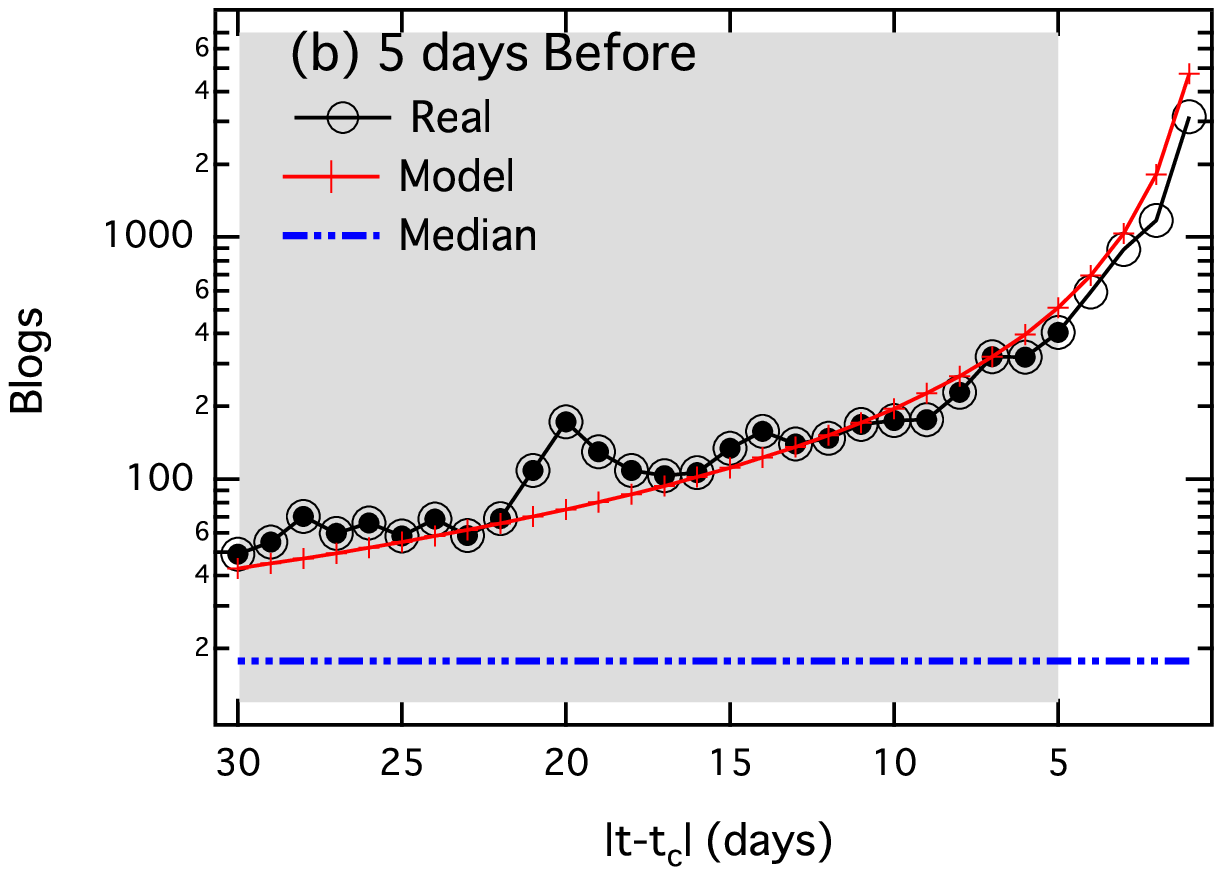}
		\caption{(Color online) Typical examples of prediction for (a) 20 days and (b) 5 days before the peak day for ``Marine Day'' in 2008 in semi-logarithmic scale. Red line indicates the prediction line, blue dashed line indicates 
median $\bar{x_j}=17$, and solid line shows the real values. 
Open circles indicate the future values and colored circles are the known values used for prediction. 
The estimation is started 85 days before $t_c$. 
Estimated values are $\alpha_j=1.78$ and $A_j=20222$ for 20 days before the peak day and 
$\alpha_j=1.38$ and $A_j=4725$ for 5 days before $t_c$. }
		\label{fig:Prediction} 
	\end{center}
\end{figure}

As an application of this study, we explore the possibility of estimating the word frequency in the near future. 
In Fig. \ref{fig:Prediction}, we show an example of prediction of blog frequency ``Marine Day'' in 2008. 
In this case,  we already have the information about the peak days to be July 21st 2008; thus, we can fix the divergence point $t_c$. 
From the data, we find that the slope period starts on April 28th, 85 days before $t_c$, as the normalized frequency continuously exceeds the median value from this day. 
In Fig.~\ref{fig:Prediction}(a), the case of prediction for 20 days before the divergence point 
using 65 data points with Eq.~(\ref{eq:model}) is shown by the red line. 
 In Fig.~\ref{fig:Prediction}(b), the case of prediction for 5 days before the peak day is shown. 
The prediction error becomes smaller for shorter prediction period as expected.
Note that a small difference in estimation of the exponent $\alpha_j$ makes a big difference near the peak; 
thus, the number of data points plays an important role in its accuracy. 
 
\section{Conclusions}\label{Sec:con} 
By analyzing a large database of Japanese blogs, we showed that the functional forms of growth and decay of word appearance that peaked on a certain day are generally approximated by power laws with the various exponents  values between -0.1 and -2.5.
The values of the power exponents depend on the category of words such as Event, Date, and News. 
In the case of Event and Date, clarification of asymmetry in the power exponents of the fore-slope and after-slope is an interesting subject for future research on collective human behavior. 
In the case of News, the power law can be observed only after the peak, and its power exponent depends on its impact. 
In the case of significant news such as the March 11th earthquake in 2011, the absolute value of the power exponent is clearly smaller than 1. 

We also checked the validity of our simple model that indicates that bloggers change their probability of posting proportional to the number of blogs and inversely proportional to the time interval from the peak. 
By checking the data of bloggers' detailed activities, we confirmed that the peaked behavior is mainly consisted of newly posted bloggers. This implies that there exists a kind of global interaction between the new comers and the keywords which makes the numbers of new comers and keywords proportional.   

In addition, these power functions can be observed also in Twitter, and it suggests that these power law behaviors are universal in social phenomena. 
An agent-based mathematical model will be used to reproduce these empirical properties of blogger activity in the near future~\cite{Yamada2}.

\begin{acknowledgments}
The authors thank Dentsu Kansai Inc. and Hottolink Inc. for useful discussions and providing the data.  
This work is partly supported by a Grant-in-Aid for 
JSPS Fellows Grant No. 219685 (H.W.) and 
Grant-in-Aid for Scientific Research No. 22656025 (M.T.)
from the Research Foundations of the Japan Society.
\end{acknowledgments}

\appendix

\section{Case without time-shift}
\label{App:NoShift}
\begin{figure}[h]
	\begin{center}
		\includegraphics[width=80mm]{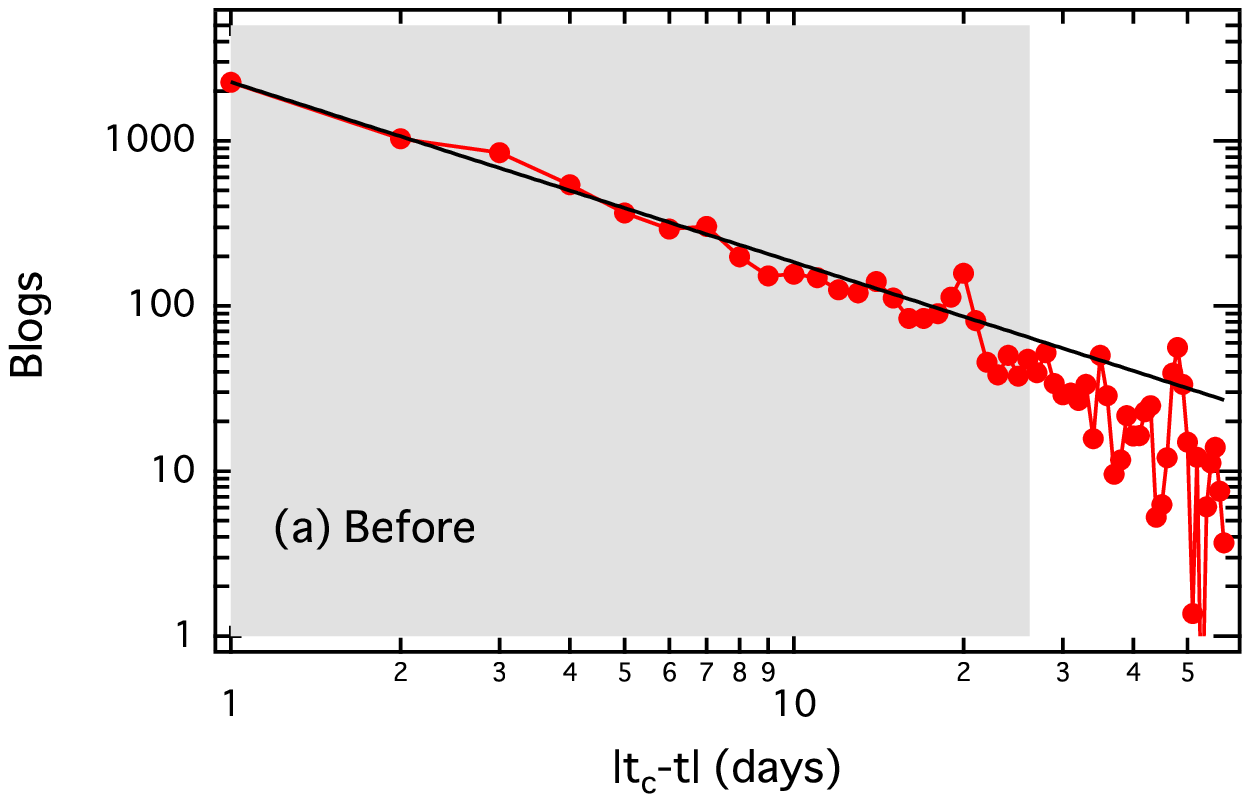}
		\includegraphics[width=80mm]{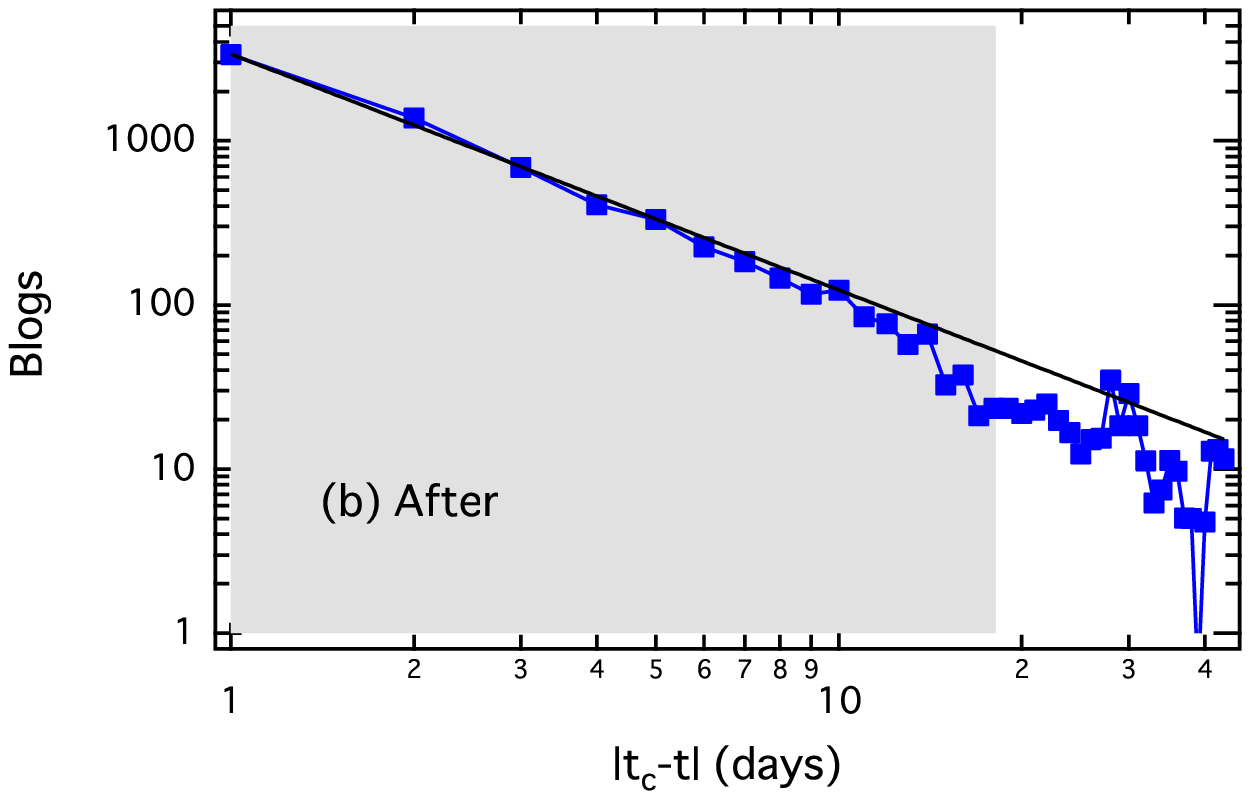}
		\caption{(Color online) Examples of data fitting without time-shift pretreatment by power laws of ``Marine Day'' in 2008 for fore-slope (a) and after-slope (b) plotted in log-log scale. For fore-slope, models are fitted by Eq.~(\ref{eq:model}) with $\alpha_j=1.10$ and $A_j=2273$ ($q=0.103$, $n=26$). For after-slope, $\alpha_j=1.44$ and $A_j=3369$ ($q=0.104$, $n=18$). The shaded area shows the interval in which the power law model is accepted with the $p$-value less than 0.1.}
		\label{fig:Fitting2}
	\end{center}
\end{figure}

\begin{figure}[h]
	\begin{center}
		\includegraphics[width=80mm]{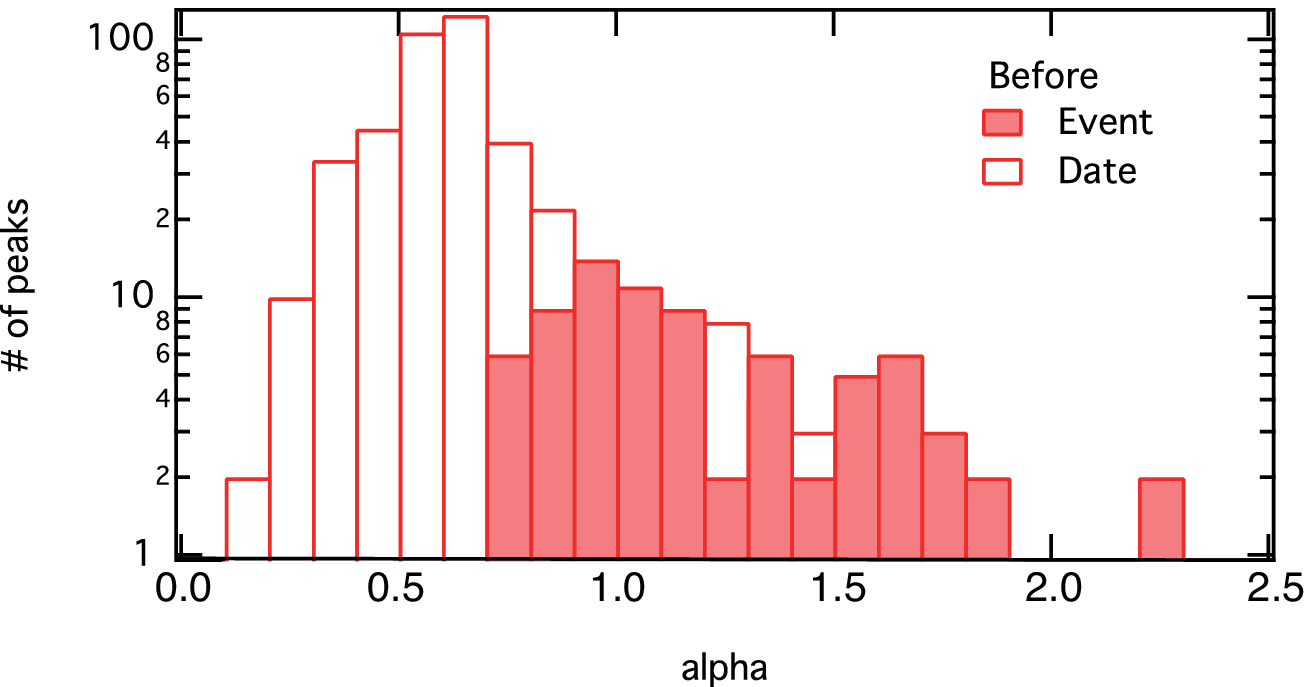}
		\includegraphics[width=80mm]{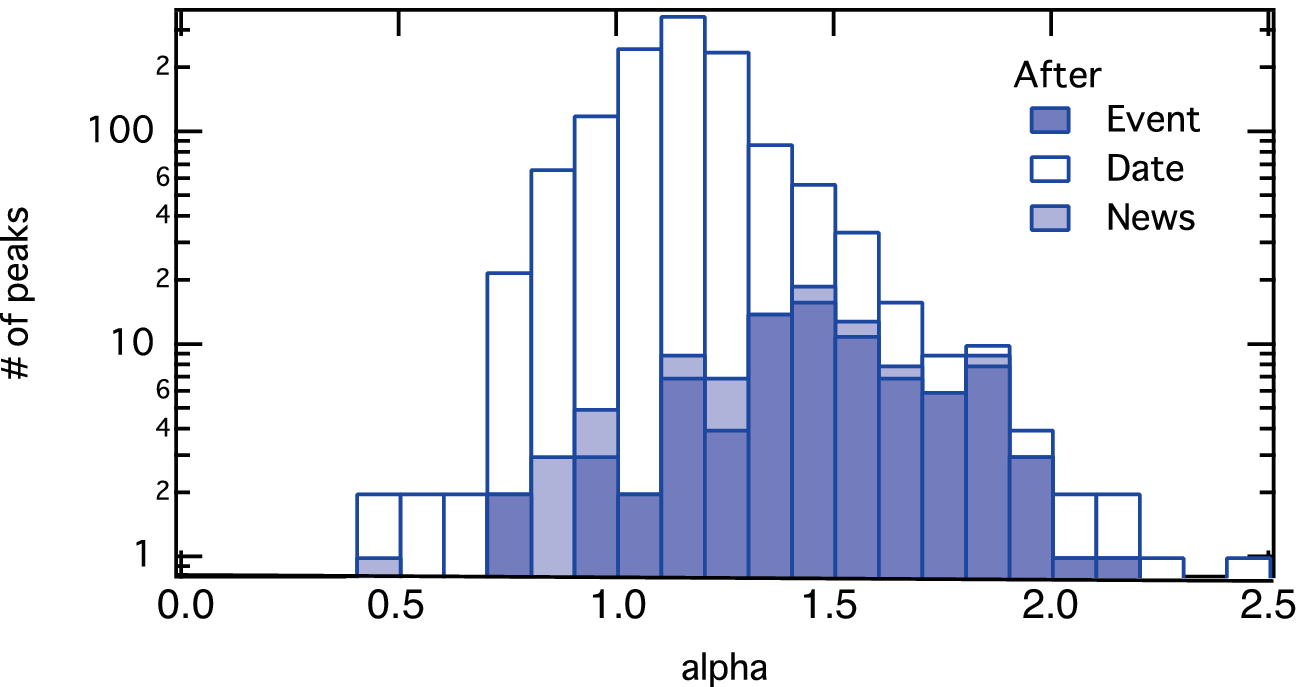}
		\caption{(Color online) Distribution of power exponent $\alpha_j$ of fore-slopes (a) and after-slopes (b) without time-shift. Mean value of $\alpha_j$ of fore-slope is 0.73$\pm$0.35 and after-slope is 1.16$\pm$0.21. Bars are colored by three categories of peaked words.}
		\label{fig:Alphas2}
	\end{center}
\end{figure}

We show the results without the time-shift; thus Eq.~(\ref{eq:ts}) with $w=1$. 
Figure~\ref{fig:Fitting2} shows a typical example of data fitting without time-shift for the word ``Marine Day'' as mentioned in Sec.~\ref{Subsec:power}. There is no major change in power exponent $\alpha_j$ for fore-slope and after-slope. However, for the value of intercept $A_j$, we can find major deviation especially for fore-slope ($A_j=3171$ with time-shift and $A_j=2273$ without time-shift).    
In Fig.~\ref{fig:Alphas2} and Table~\ref{tab:Alpha2}, we summarize the whole samples.  
 \begin{table}
	\caption{Mean values of power exponent $\alpha_j$ with standard deviations and medians of slope days $n$ in case without time-shift procedure which introduced in Sec.~\ref{Subsec:pre}.}
	\label{tab:Alpha2}
	\begin{center}
		\begin{tabular}{ccccc} \hline
			 &  & $\alpha_j$ & $n$ (days) & \# samples\\ \hline \hline
			Event & Before &1.21 $\pm$ 0.38 & 15.5 & 80 \\ 
			 & After &  1.48 $\pm$ 0.28 & 16  & 85 \\ 
			Date & Before & 0.64 $\pm$ 0.25 & 11.5  & 418 \\ 
			 & After & 1.14 $\pm$ 0.18 & 22  &1176  \\ 
			News & After & 1.21 $\pm$ 0.35  & 11  & 18 \\ \hline
			All & Before & 0.73 $\pm$ 0.35 & 12 & 498  \\
			 & After & 1.16 $\pm$ 0.21 & 21  &1279 \\ \hline \hline
		\end{tabular}
	\end{center}
\end{table}

\section{Modified Random Diffusion Model}
\label{App:ERD}
\begin{figure}
	\begin{center}
		\includegraphics[width=80mm]{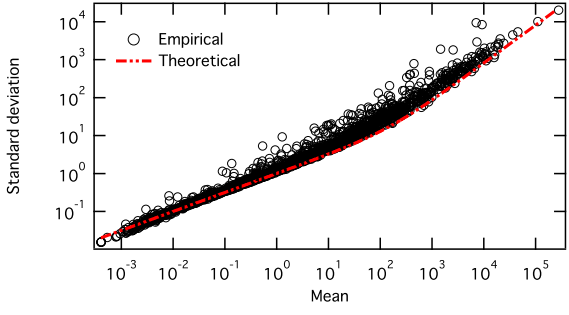}
		\caption{(Color online) Relationship between mean and standard deviation of word frequency in the blogosphere.  Empirical results of 1771 adjectives and theoretical result of Eq.~(\ref{eq:App13}) are duplicated in the figure.}
		\label{fig:AveSta}
	\end{center}
\end{figure}

We introduce a modified random diffusion model, which is used in Eq.~(\ref{eq:RD}). 
The random diffusion model was originally introduced to describe diffusion properties of random walkers on a given network  \cite{Barabasi2,Meloni}, 
and two of the authors (Y.S. and M. T.)  have modified  the model to be applicable to the fluctuations in word appearance in the blogosphere  \cite{Sano}. 
In our modified random diffusion model, we assume that there are two states, 
active and non-active for each blogger, and the number of active bloggers fluctuates randomly each day. 
Each active blogger randomly decides to post a blog including the $j$-th word. 
There is a key parameter in this stochastic process; the share of the $j$-th word $c_j$ is defined by the following equation 
\begin{equation}
	c_j=\frac{\langle x_j \rangle}{\langle X \rangle},
	\label{eq:App1}
\end{equation}
where $x_j(t)$ is the number of blog entries including the $j$-th word on the $t$-th day. $X(t)$ is the number of active bloggers on the $t$-th day 
and the brackets show the mean over all instances. 
We assume that the number of active bloggers $X(t)$, $X(t) \ge 0$
fluctuates randomly following an independent probability density distribution $\phi(X)$ with finite moments. 
Probability of posting $x_j$ entries is calculated using a Poisson distribution with the mean number $c_jX$ given as follows
\begin{equation}
	P(x_j|c_j)=\int_0^{\infty}{\phi(X)\exp\left(-c_jX\right)\frac{\left(c_j X\right)^{x_j}}{x_j!}dX}. 
	\label{eq:App4}
\end{equation}
When $\langle x_j \rangle$ is small,  
a Poisson distribution is approximated by a Bernoulli distribution that assumes $x_j = 0$ with a probability $1 - c_j X$, and 
$x_j = 1$ with a probability $c_j X $.  
Thus, we have the following evaluations for an arbitrary distribution of $\phi(X)$.
\begin{eqnarray}
	P(x_j=0|c_j) & \simeq &\int_0^{\infty}{\phi(X) \left( 1- c_j X \right)dX} \nonumber \\ 
	& \simeq & 1-c_j\langle X \rangle, \nonumber \\
	P(x_j=1|c_j) & \simeq &\int_0^{\infty}{\phi(X) \left( c_j X \right)dX} \nonumber \\ 
	& \simeq & c_j\langle X \rangle.
	\label{eq:App4.2}
\end{eqnarray}
For  $\langle x_j \rangle \approx 2$, $P(x_j\ge 2 | c_j) \approx 0$, 
thereby $P(x_j|c_j)$ is approximated by the Poisson distribution with both the mean and the variance given by $c_j \langle X \rangle$. \\
For $\langle x_j \rangle\gg1$, the Poisson distribution can be approximated by a normal distribution, 
\begin{equation}
	P(x_j|c_j)\simeq \int_0^{\infty}{\phi(X)\frac{1}{\sqrt{2\pi c_j X}}\exp{\left[-\frac{(x_j-c_jX)^2}{2c_j X }\right]}dX}.
	\label{eq:App5}
\end{equation}
By introducing a new variable $y_j=\frac{x_j}{c_j \langle X \rangle}$, Eq. (\ref{eq:App5}) becomes
\begin{equation}
	P(y_j|c_j)\simeq \int_0^{\infty}{\phi(X)\frac{1}{\sqrt{2\pi \left (\frac{X}{c_j {\langle X \rangle}^2}\right)}}\exp{\left[-\frac{(y_j-\frac{X}{\langle X \rangle})^2}{2\left(\frac{X}{c_j { \langle X \rangle}^2}\right)}\right]}dX}.
	\label{eq:App6}
\end{equation}
When $\langle x_j \rangle =c_j \langle X \rangle \gg 1$, the weight function in the 
integral can be approximated by Dirac's delta function as
\begin{equation}
	P(y_j|c_j)\simeq \int_0^{\infty}{\phi(X)\delta\left(y_j- \frac{X}{\langle X \rangle}\right)dX}.
	\label{eq:App7}
\end{equation}
Therefore, we have the following simple evaluation, for $x_j$, 
\begin{equation}
	P(x_j|c_j)\simeq \frac{1}{c_j}\phi\left(\frac{x_j}{c_j}\right).
	\label{eq:App8}
\end{equation}
Calculating the first and second moments of $P(x_j|c_j)$, we now have the general results
\begin{eqnarray}
	\langle x_j \rangle & = & \int_{0}^{\infty}{x_jP(x_j|c_j)}dx_j \nonumber \\ 
	&\simeq&\int_{0}^{\infty}{x_j \frac{1}{c_j} \phi \left( \frac{x_j}{c_j} \right)}dx_j=c_j\langle X \rangle, \\ 
	\langle x_j^2 \rangle & = & \int_{0}^{\infty}{x_j^2P(x_j|c_j)}dx_j   \nonumber \\
	&\simeq & \int_{0}^{\infty}{x_j^2 \frac{1}{c_j}  \phi \left( \frac{x_j}{c_j}\right )}dx_j = {c_j}^2{\langle X^2\rangle}. \\
	\nonumber
	\label{eq:App10}
\end{eqnarray}
From these results, the standard deviation $\sigma_j=\sqrt{\langle x_j^2 \rangle - {\langle x_j \rangle}^2}$ can be expressed as 
\begin{equation}
	\sigma_j \simeq \sqrt{{c_j}^2 \left ( \langle X^2 \rangle- {\langle X \rangle}^2 \right)}. 	
	\label{eq:App11}
\end{equation}
By correlating both results Eqs.~(\ref{eq:App4.2}) and (\ref{eq:App11}), we can get the following relation; 
\begin{equation}
	\sigma_j \simeq \sqrt{c_j\langle X \rangle + {c_j}^2  {\langle X^2 \rangle}_c}, 
	\label{eq:App12}
\end{equation}
where ${\langle X^2 \rangle}_c$ denotes the second order cumulant. By using $\langle x_j \rangle =c_j \langle X \rangle$, we rewrite Eq.~(\ref{eq:App12}) into
\begin{equation}
	\sigma_j \simeq \sqrt{\langle x_j \rangle + \left (1 + \frac{{\langle X^2 \rangle}_c}{{\langle X \rangle}^2} \langle x_j \rangle \right) }. 
	\label{eq:App13}
\end{equation}
Figure~\ref{fig:AveSta} shows empirical results using 1771 adjectives and Eq.~(\ref{eq:App13}) with $\frac{{\langle X^2 \rangle}_c}{{\langle X \rangle}^2}=0.08$.

\end{document}